\documentclass[%
 reprint,
 amsmath,amssymb,
 aps,
]{revtex4-2}

\usepackage{xr}
\makeatletter
\newcommand*{\addFileDependency}[1]{
  \typeout{(#1)}
  \@addtofilelist{#1}
  \IfFileExists{#1}{}{\typeout{No file #1.}}
}
\makeatother

\newcommand*{\myexternaldocument}[1]{%
    \externaldocument{#1}%
    \addFileDependency{#1.tex}%
    \addFileDependency{#1.aux}%
}

\myexternaldocument{Supplementary_materials_2.0}

\usepackage{graphicx}
\usepackage{dcolumn}
\usepackage{bm}

\begin{document}

\preprint{APS/123-QED}

\title{Ups and downs: Copepods reverse the near-body flow to cruise in the water column}

\author{Nils B. Tack}
 \affiliation{ School of Engineering, Brown University, 345 Brook St., Providence, RI 02912, USA.}
\author{Sara Oliveira Santos}%
 \affiliation{ School of Engineering, Brown University, 345 Brook St., Providence, RI 02912, USA.}
\author{Brad J. Gemmell}
 \affiliation{Department of Integrative Biology, University of South Florida, 4202 East Fowler Ave., Tampa, FL 33620, USA.}
\author{Monica M. Wilhelmus}%
 \email{mmwilhelmus@brown.edu}
 \affiliation{ School of Engineering, Brown University, 345 Brook St., Providence, RI 02912, USA.}
 \email{mmwilhelmus@brown.edu}


\date{\today}

\begin{abstract}
Copepods participate in large-scale diel vertical migrations (DVM) as primary consumers in marine ecosystems. Given that they are negatively buoyant, gravity facilitates their downward cruising but impedes their upward relocation. In principle, overcoming this retarding force using drag-based propulsion during upward swimming incurs extra energy costs that neutrally buoyant or gravity-assisted downward swimmers fundamentally avoid. This added power requirement seemingly renders upward swimming disproportionally less efficient than downward swimming. However, we found
a compensatory mechanism enhancing thrust in upward-swimming copepods. Using experimentally derived velocity and pressure fields, we observed that copepods pull water to the anterior, generating sub-ambient pressure gradients when cruising upwards, thereby inducing an upstream suction force to complement the thrust produced by the legs.
Contrary to expectations that drag forces always dominate the leg recovery phase, the results show that the upstream suction generates net thrust for about a third of the recovery stroke. Such a suction-thrust mechanism promotes cost-effective swimming in larger animals like fish and jellyfish. However, it has never been reported at small scales. We compare these results with downward-swimming copepods that push rather than pull on water and demonstrate they do not benefit from thrust-enhancing suction effects during the recovery stroke. Instead, downward cruisers leverage gravitational effects to swim faster. `Puller' and `pusher' behaviors are driven by alterations of the leg kinematics, indicating a response to the body orientation rather than a fixed organism trait. These results offer insights into an essential swimming mechanism that can inform us about the role of mesozooplankton in driving biogenic hydrodynamic transport and its effects on marine biogeochemistry. 
\end{abstract}

\maketitle


\section{\label{sec:level1}Introduction}
Copepods are one of the most abundant mesozooplankton in marine ecosystems \cite{Humes1994}. They are found in a broad range of marine environments spanning coastal waters and open ocean while taking part in large-scale mass movements known as diel vertical migrations (DVM), which are central to their ecological role as primary consumers \cite{Gauld1953,Roe1972,conroy2020zooplankton}. They cover vertical distances of up to several hundred meters to forage in shallow waters \cite{Bianchi2013, Bianchi2016}. This migration is mainly controlled by light, but can also be influenced by predation pressure, food availability and advection risk \cite{benoit2021vertical, benoit2019forage, huntley1982effects, sato2024diel}. Vertically migrating mesozooplankton have a complex role in biogeochemical cycling, notably the biological pump regulating carbon transfer and transport from primary producers at the surface into the deep ocean. By shunting carbon via the release of surface-derived metabolites, diel vertical migrators can increase particulate organic carbon (POC) transfer efficiency in the upper ocean and fuel primary production in nutrient-limited areas \cite{Steinberg2000} (reviewed in \cite{Cavan2019, Steinberg2017}). Considering translocating organism-level functions alone (such as grazing and excretion), DVM-mediated transport of organic matter has been estimated to contribute 14--40$\%$ of the global export flux, along with redistributing oxygen profiles \cite{Bianchi2013,Aumont2018,Archibald2019}. Vertical migrations have thus been hypothesized to potentially have important effects on the local-to-global biogeochemistry of the ocean \cite{Wilhelmus2014, Wilhelmus2019, houghton2019alleviation,Siegel2023}. However, because of the complexity of the pathways that link the upper ocean to the interior and the widely different regulatory mechanisms (i.e., biotic and abiotic factors), it is challenging to quantify the impact diel vertical migrators have on the global carbon cycle \cite{Siegel2023,Hansen2016,Gorgues2019}. The net contribution of induced hydrodynamic transport is critical to constrain the role of mesozooplankton in the biological pump properly. While the mixing by an individual swimmer is found to be small \cite{Visser2007,Noss2014}, laboratory measurements showed swarm-induced flows producing hydrodynamic instabilities capable of inducing large-scale mixing \cite{Wilhelmus2014,Houghton2018,Wickramarathna2014,Noss2012}. This is because hydrodynamic interactions from individual contributions in the collective vertical migration of swarms produce a spatially coherent wake in the opposite direction of the migration \cite{ouillon2020active}. To contextualize the relevance of these effects in oceanic ecosystems further, it is essential first to understand the swimming characteristics and hydromechanical mechanisms enabling the vertical relocation of individual plankton.\\

In negatively buoyant calanoid copepods, the force exerted by gravity is critical during vertical movements \cite{Clarke1934,Haury1976,Jiang2023}. On the one hand, cruisers swim faster going downwards by virtue of swimming in the direction of sinking, whereby the terminal velocity contributes to increasing performance \cite{Jiang2002}. On the other hand, to achieve propulsion, hovering and upward swimming copepods must at least generate forces that counterbalance their excess weight (equal in magnitude but opposite to gravity) \cite{Jiang2004}. The anchoring effect induced by excess weight enables the production of a stronger anterior velocity gradient, pulling water toward the copepod \cite{Strickler1982, Emlet1985}. The implications of the relative (excess) weight acting as an anchor to pull harder on the surrounding water indicate that copepods need to generate more power to overcome gravity, likely increasing the cost of transport (COT) \cite{Marshall1972,Jiang2023}. This additional power demand creates a noticeable disparity in efficiency between upward and downward swimming.\\ 

Daily migrations are thought to be energetically expensive. Based on the observed duration and amplitude of DVMs of calanoid copepods, the physiological cost is estimated to range between 13$\%$ and 120$\%$ of the basal metabolic rate (reviewed in \cite{Mauchline1998}). This wide range depends on the species, estimated swimming speed, and swimming mode. How copepods perform extensive cruising despite energetic limitations has been thus far evaluated from an ecological standpoint, but hydrodynamic effects need to be considered. For instance, numerical simulations suggest that producing a strong, pulling anterior flow can enhance the prey capture volume to offset the energy budget of migration and satisfy the energy need of upward cruising \cite{Jiang2002}. This argument is partly supported by the fact that grazing motivates upward swimming, especially during DVMs \cite{Gauld1953}. However, in this context, compensating for, rather than mitigating energy losses makes vertical swimming highly dependent upon unpredictable external factors, such as food density and abundance, with potentially undesired consequences.\\

From a bio-fluids perspective, an alternative explanation for the role of the strong velocity gradient produced by upward cruisers stems from how some animals, including fish and jellyfish, pull themselves through the water using suction thrust \cite{Gemmell2015,Colin2012,Gemmell2016}. By accelerating the surrounding fluid -- such as when pulling on water -- counter-rotating vortices form, at the interface of which a high-velocity, low-pressure region exists \cite{Dabiri2020,Colin2012}. The reciprocal action of this local low pressure anterior to the body generates a suction force in the swimming direction, contributing to thrust \cite{Gemmell2015}. This pull-thrust mechanism promotes economical swimming and enhances performance in fish \cite{Gemmell2015,Tack2021}, jellyfish \cite{Colin2012,Dabiri2020}, and ctenophores \cite{Colin2020}. 
This is because the inertia carried by the persisting induced flow can be harnessed at no additional cost. This potentially offers negatively buoyant copepods a solution to overcome their excess weight at reduced cost when cruising vertically. We hypothesize that by metachronally beating the cephalic appendages in certain ways, upward-swimming copepods can generate counter-rotating vortices that advect water anteriorly and around their body to harness similar suction effects to counter gravity. In contrast, we expect downward swimmers to achieve greater cruising speeds but no longer benefit from their relative weight to generate a strong pulling force.\\

The goal of this study is to evaluate the impact on thrust production and swimming ability. Do copepods actively modulate the flow around their body in response to orientation when cruising in the water column? Our results provide new insights into important hydrodynamic mechanisms at the organismal level, whose cumulative effects in large aggregations during DVMs can potentially impact the vertical distribution of marine biogeochemical properties \cite{Wilhelmus2014,Wilhelmus2019, houghton2019alleviation, ouillon2020active}.\\

\section{\label{sec:level1}Methods}
 Adult copepods \textit{T. longicornis} (prosome length = 0.6--0.8 mm) were collected from a pier in Woods Hole, MA, USA, and acclimated overnight at room temperature (approx. 21$^{\circ}$C). Observations were made in a small filming vessel (10 ml) containing a dilute suspension (3 to 5 copepods) such that the flow field of the observed local organism was not affected by that of other swimmers. Bright-field 2D-Particle Image Velocimetry (PIV) was performed following methods by \cite{Gemmell2014}. The water was seeded with heat-killed microalgae \textit{Nannochloropsis oculata} ($\approx 2 \mu$m in diameter), backlit by a 150 W fiber optic illuminator (Fisher Scientific) coupled with a collimating lens to visualize the flow. The light source did not induce phototaxis. Free-swimming copepods were recorded dorsoventrally using a high-speed digital video camera (Fastcam Mini WX 100; Photron, Tokyo, Japan) at 2000 fps ($1024\times1024$ pixels). Despite using volume illumination, the camera was equipped with a high-magnification optical set-up of a narrow depth-of-focus (DoF approx. 127 $\mu$m, see supplementary materials). In total, dozens of videos were recorded, but it must be noted how challenging it is to capture a copepod in such a narrow 2D field for a significant period in an open 3D filming vessel away from walls. We selected two sequences of upwards and downwards swimming when the copepods are fully in view for several consecutive leg beats.
 Fluid velocity vectors were calculated using the DaVis 10 software package (LaVision, Göttingen, Germany). Image pairs were analyzed with three passes of overlapping interrogation windows (75\%) of decreasing size of $48\times48$ pixels to $32\times32$ pixels. Manually masking the body of the copepods before image interrogation confirmed the absence of surface artifacts in the PIV measurements. All frames were used for analysis, yielding a separation between frames ($\Delta$\textit{t}) of $5\times10^{-4}$ s.\\

\begin{figure}
    \centering
    \includegraphics[width=0.5\textwidth]{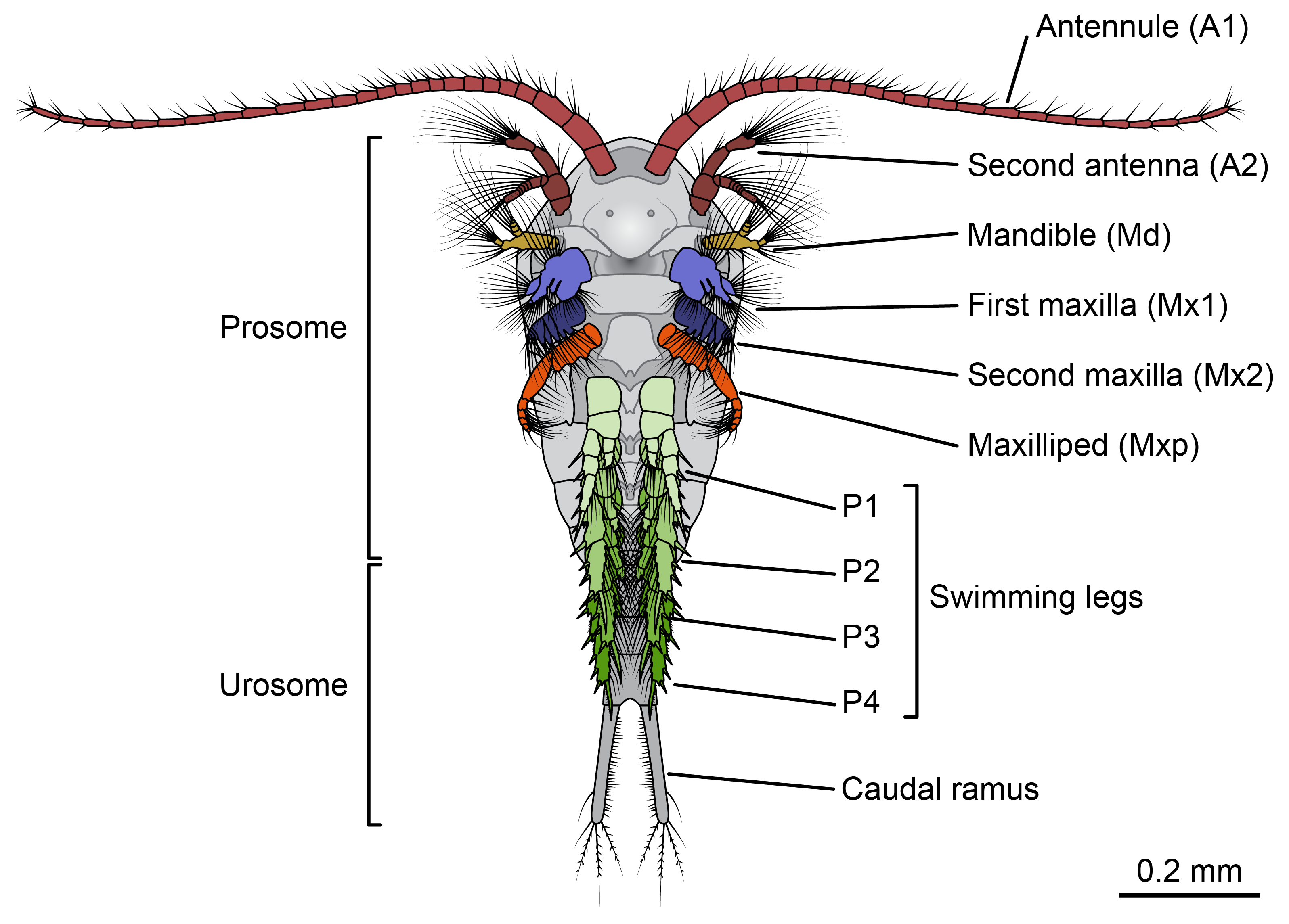}
    \caption{\textit{Temora longicornis} external morphology (ventral view). This view highlights the location of the five cephalic appendages employed during swimming in our experiments (A2, Md, Mx1, Mx2, and Mxp). Other swimming legs (P1--P4) are generally employed during fast swimming. Copepod length was measured as the prosome length. 
    }
\label{fig:Diagram}
\end{figure}

Morphometrics and swimming kinematics measurements were performed from the scaled PIV videos for each copepod using ImageJ software. The locomotor classification was determined for each animal after computing the velocity fields and was based on the direction of the dominant anterior flow. Motions of the second antennae (A2), the dominant propulsive cephalic appendages, were measured in degrees relative to the swimming direction and normalized to 180\textdegree (corresponding to the lateral halves of the body). Kinematics parameters were averaged for several consecutive leg beats from the beginning of the video sequence to either the end of the sequence or when the copepod leaves the field of view (see Table ~\ref{tab:table1}). Note that other cephalic appendages, including the mandibles (Md), first (Mx1) and second (Mx2) maxillae, and maxillipeds (Mxp) (see Fig.~\ref{fig:Diagram}), were also beating metachronaly during swimming. The antennules (A1)  were not involved in locomotion.\\

Pressure fields around the body of the copepods were computed using the Queen 2.0 package for Matlab \cite{Dabiri2014, Lucas2017}. Given the sensitivity of this calculation to standard PIV errors at the fluid-solid interface, the copepod body shapes were manually masked before image interrogation, ensuring the absence of surface artifacts in the PIV measurements. While two-dimensional, this approach accurately estimates the pattern, timing, and magnitude of pressure fields around fish-like swimmers \cite{Dabiri2014} and zooplankton \cite{Colin2020}. Note that the final pressure estimates are relative to a zero reference pressure corresponding to the surrounding ambient pressure (gauge pressure).\\

Forces were computed from the pressure fields to quantify the contribution of positive and negative pressures to thrust and drag \cite{Lucas2017}. Force magnitude was calculated per unit depth (because PIV data were 2D) as the product of the length of each segment between points making up the outline of the copepod, the pressure along each segment, and the unit vector normal to each segment, giving units of Newtons per meter. Thrust and drag are the axial components of the forces in the swimming direction, $u$ (see supplementary materials). Forces were classified as pull and push forces when arising from sub-ambient pressures (negative relative to ambient) and above-ambient pressures (positive), respectively. The forces produced over time were averaged for several consecutive beat cycles (see Table ~\ref{tab:table1}).\\

\section{\label{sec:level1}Results}
In total, we studied four freely swimming copepods; two animals swimming upward ($0 < u < \pi$) and two specimens swimming downward ($-\pi < u < 0$) (see Table~\ref{tab:table1}). Observations in which the copepods moved out of the focal plane or exhibited undesirable behaviors, such as jumping, were excluded from the analysis. While the upward-swimming copepods swam nearly vertically ($u$ was less than 3\textdegree offset from the vertical laboratory frame of reference), one of the downward-swimming copepods displayed a strong horizontal component (see Table ~\ref{tab:table1}, Supplementary S\ref{fig:swimDirection}). We determined this had little to no effect on their behavior because induced drag is the only force opposing their motion. We also confirmed the display of the cruising rather than the feeding behavior (or a combination thereof) with the absence of the characteristic kinematics and flow features commonly associated with the production of a feeding current \cite{Strickler1982,Strickler1984,Cannon1928} (see supplementary materials).\\ 

Both upward swimmers were slower than the downward cruisers. Upward swimmers had speeds $U$ = 2.37 and 6.99 BL s$^{-1}$ while downward cruisers swam at $U$ = 12.54 and 10.78 BL s$^{-1}$ (Table ~\ref{tab:table1}). The terminal sinking speed $\omega_s$ of the copepods was not determined experimentally, but data from the literature indicate $\omega_s$ =  2.5 to 2.9 mm s$^{-1}$ for another calanoid copepod species, \textit{T. longicornis} \cite{Apstein1910,Tiselius1990}. We used a modified Stokes' law for irregularly shaped particles to confirm that our experimental animals had sinking speeds within this range. This method provides a robust estimate of the sinking rate to determine the terminal sinking speed of copepod carcasses \cite{Elliot2010}. We found $\omega_s$ = 2.68 $\pm$ 0.05 mm s$^{-1}$, equivalent to 3.82 $\pm$ 0.41 BL s$^{-1}$ (see supplementary materials).The drag coefficient may fluctuate in self-propelled organisms, thus giving only estimates of $\omega_s$ \cite{Jiang2023}. Nonetheless, knowing $\omega_s$ is helpful in showing that the swimming speed achieved by downward swimmers is not solely the result of their sinking rate. Indeed, $\omega_s$ of downward-swimming copepods is 30.5 to 35.4$\%$ of $U$.\\

In all cases, cruising was achieved using the same swimming mode -- by metachronally beating the cephalic appendages. However, swimming upward induced `breaststroke' kinematics consisting of the cephalic appendages extending anteriorly at the beginning of the power stroke (normalized start angle = 0.26$\pm$0.03 relative to $u$) and laterally at the end of this phase (normalized end angle = 0.55$\pm$0.06). In contrast, downward swimmers initiated their power stroke medially (normalized start angle = 0.52$\pm$0.05) and terminated it posteriorly (normalized end angle = 0.75$\pm$0.04) (Table ~\ref{tab:table1}, Supplementary S\ref{fig:legPosition}). Other swimming parameters remained unaffected; upward and downward swimmers had comparable leg beat frequencies, with the former averaging 58.3$\pm$0.6 s$^{-1}$ and the latter 59.5$\pm$1.8 s$^{-1}$.\\

\begin{figure}
    \centering
    \includegraphics{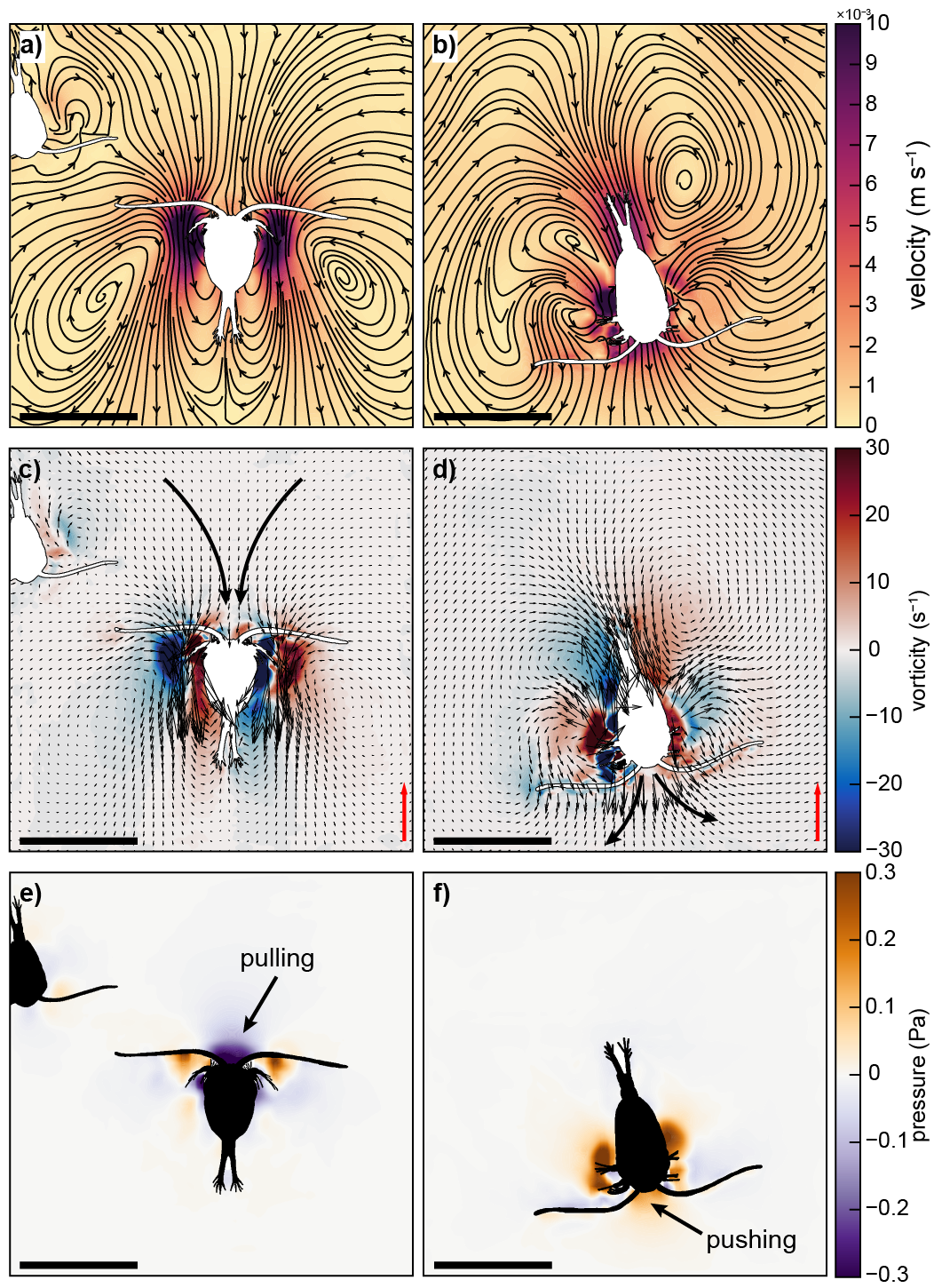}
    \caption{\label{fig:VeloVortPress} Instantaneous velocity, vorticity and pressure fields for upward and downward swimming copepods. Velocity magnitude and streamlines (in a
lab-reference frame) of a slow, ascending (a), and a fast downward swimming copepod (b). Upward swimmers pull the water toward them, while downward swimmers push the water away. We classify copepods displaying these characteristics as 'pullers' and 'pushers', respectively. (c,d) Vorticity fields show the different flow characteristics of pullers and pushers, respectively. Thick arrows indicate the dominant flow produced in front of the copepods. The red scale arrow in (c,d) indicates $1\times10^{-2}$ m s$^{-1}$. Every three vectors were plotted for clarity. (e) Experimentally derived pressure fields show that copepods drop the pressure in front of them by pulling water in when ascending. (f) In contrast, downward swimming 'pushers' generate a high-pressure area in front of them. The black scale bar indicates 1 mm in all the panels.
    }
\end{figure}

\begin{figure}
    \centering
    \includegraphics[width=0.5\textwidth]{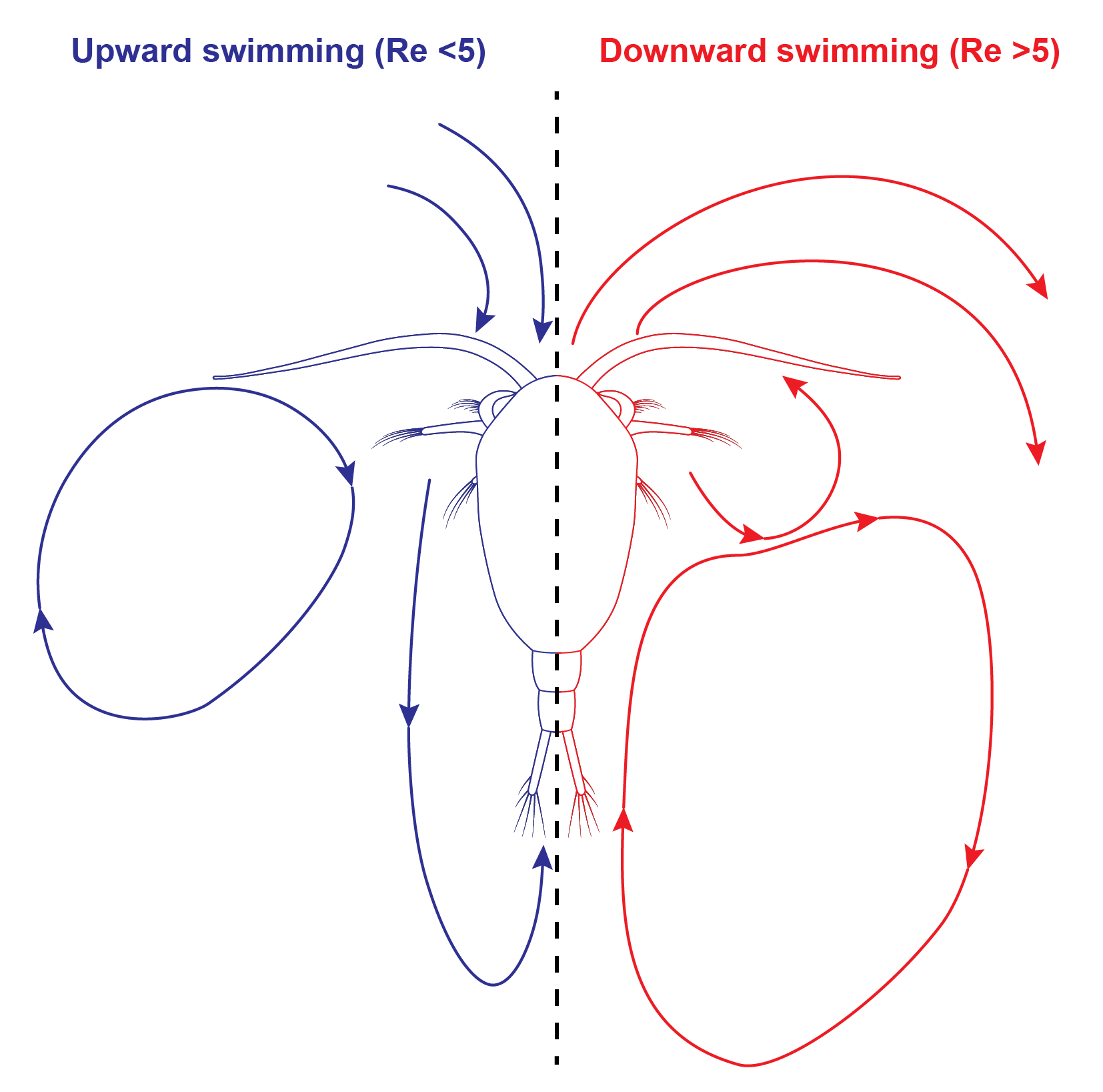}
    \caption{Schematics of the near-field flow produced by downward and upward swimming copepods. Upward swimming copepods advect a large funnel-like volume of water anteriorly that is expelled posteriorly in momentum jets (left). They produce two vortices on each side of the body; one large vortex adjacent to the prosome and another laterally compressed vortex extending posteriorly along the urosome. Downward swimmers push the water in front of them and also form two observable vortices (right). One small vortex surrounds all the cephalic appendages and a much larger counter-rotating vortex is located posteriorly and extends far behind the urosome .}
\label{fig:Schematics}
\end{figure}

The contrasting kinematics and orientations discussed above caused two distinct near-field flow structures (Figs~\ref{fig:VeloVortPress},\ref{fig:Schematics}). Slow, upward-swimming copepods acted as `pullers', entraining the water toward them, as seen with the flow converging anteriorly (Figs~\ref{fig:VeloVortPress}(a),(c),\ref{fig:Schematics}). The pulling action of the cephalic appendages caused a local drop in pressure anteriorly (Fig.~\ref{fig:VeloVortPress}(e)). In contrast, downward-swimming animals behaved as `pushers', as evidenced by anterior fluid flow displacement in the swimming direction and the subsequent lateral deflection of the water (Figs~\ref{fig:VeloVortPress}(b),(d),\ref{fig:Schematics}). This resulted in a local increase in pressure in front of the copepods (Fig.~\ref{fig:VeloVortPress}(f)). The shape and distribution of the induced vortices around the body also differed substantially. Upward swimmers produced vortex pairs on each side of the body; one large vortex lateral to the prosome (body) and another laterally compressed vortex extending posteriorly along the urosome (tail) (see Fig.~\ref{fig:Schematics}). Fast, downward-cruising specimens also produced two vortices: one small vortex surrounding all the cephalic appendages and a much larger counter-rotating vortex located posteriorly and extending far behind the urosome (Fig.~\ref{fig:VeloVortPress}(b),(d)). Note that the larger vortex forming in the upward and downward swimming cases have opposite signs.\\

\begin{figure}[h!]
    \centering
    \includegraphics{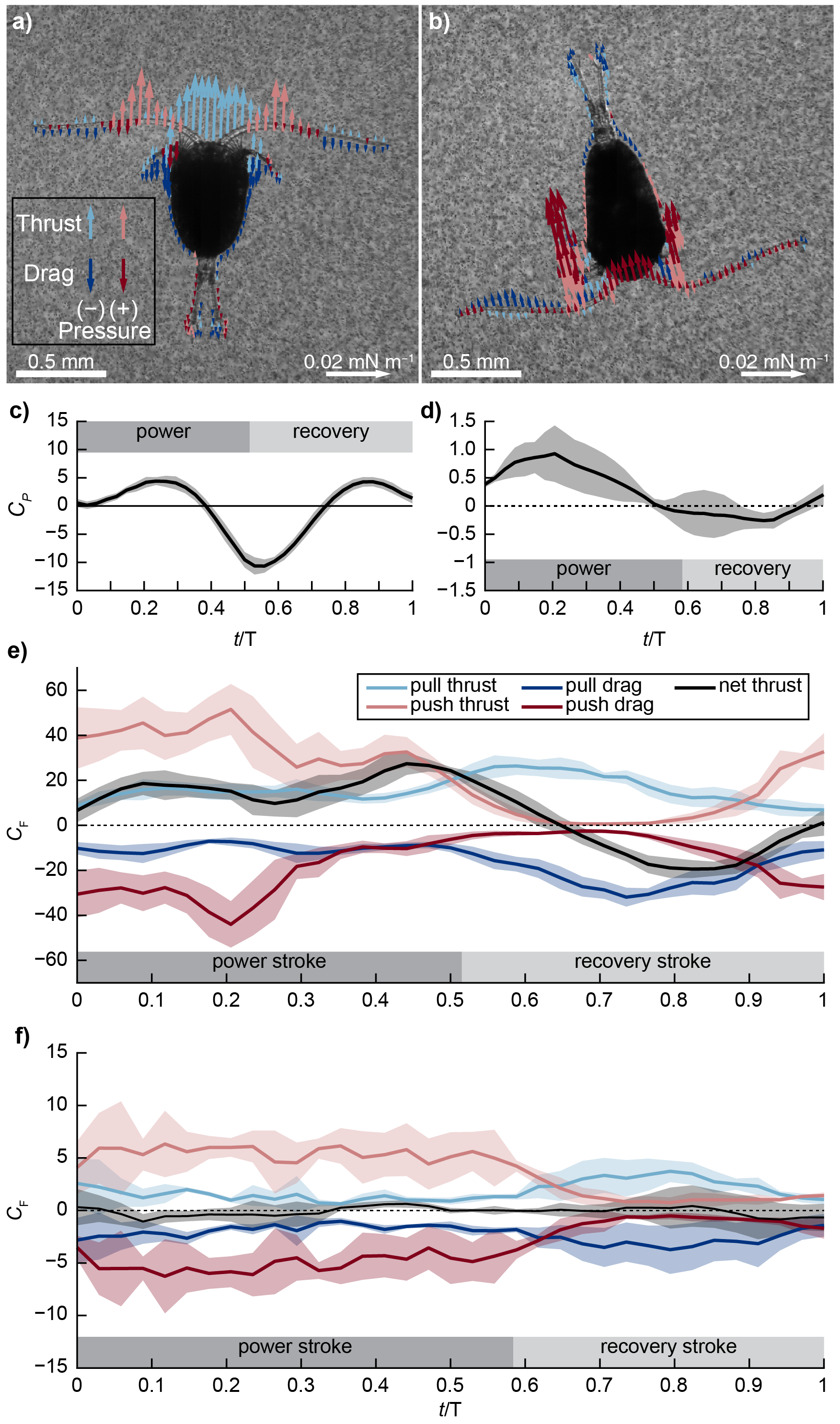}
    \caption{Instantaneous forces produced during swimming for representative upward and downward swimmers. Instantaneous force vectors at the end of the power stroke for a representative slow, upward swimming (a) and a fast, downward swimming copepod (b). Note that only the axial component is plotted to show the net contribution to thrust and drag acting in the swimming direction. Every two vectors are plotted for clarity. The sign of the pressure coefficient in front of upward swimmers (c) changes twice during a beat cycle, while it changes only once in downward cruisers (d). Downward swimmers generate strong sub-ambient pressures at the beginning of the recovery stroke. (e) Mean instantaneous force coefficients ($C_{F}$) for the upward-swimming copepod depicted in (a) and (c). (f) Mean instantaneous $C_{F}$ for the downward-swimming copepod depicted in (b) and (d). Solid lines in (c-f) indicate the mean for 10 and 3 leg beat cycles for each representative upward and downward swimmer, respectively, and shading shows the standard deviation.
    }
    \label{fig:Forces}
\end{figure}

In the case of upward swimming copepods, the anterior pressure fluctuations coincided with oscillations in the net thrust. Downward swimmers, however, generated nearly constant net thrust (Fig.~\ref{fig:Forces}). In general, the net thrust oscillated around zero over a complete beat cycle. This is expected because, by definition, during steady swimming, the thrust and drag forces – and the gravitational force in upward swimmers – are balanced, resulting in no net time-average acceleration. Upward swimmers generated positive net thrust throughout the entirety of the power stroke and the initial phase of the recovery stroke. Net drag was produced during the remainder of the leg recovery phase (Fig.~\ref{fig:Forces}(e)).\\

 Overall, the balance of forces (i.e., thrust and drag) was comparable between the two swimming cases during the power stroke. Dominant push forces were relatively constant during the first half of the power stroke and gradually dropped before initiating the recovery stroke (Fig.~\ref{fig:Forces}(e),(f)). However, upward swimmers also produced positive net thrust at the beginning of the recovery stroke due to strong pull forces at the front of the body (Fig.~\ref{fig:Forces}(a)). Net thrust was produced during 27.2 and 31.2\% of the duration of the recovery stroke of copepods 8 and 5, respectively. This is consistent with the pulling behavior and anterior local pressure drop described above (Figs~\ref{fig:VeloVortPress},\ref{fig:Forces}(c)). Drag eventually dominated later in the recovery stroke due to the vortex developing along the prosome that entrained water posteriorly to the body, thus increasing pull drag. These effects were not present in the fast cruisers, as shown by the overall balance between these forces during the recovery stroke. In this case, at the beginning of the recovery stroke, the anterior flow was mostly dominated by positive pressure gradients, which generated drag (Figs~\ref{fig:VeloVortPress}(f),\ref{fig:Forces}(b),(d)). Note that the upward-swimming copepods have greater force coefficients than the downward swimmers. This is because the raw force magnitudes of both cases are equivalent, but upward copepods are slower. Thus, the latter would produce proportionally more force and potentially more power to achieve the same swimming speed as downward cruisers.\\

\begin{table*}[t]
\caption{\label{tab:table1} Copepod morphometrics and swimming parameters. Parameters measured for each copepod over several individual consecutive appendage beats are reported as mean$\pm$standard deviation.}
\begin{ruledtabular}
\begin{tabular}{ccccc}
Copepod ID & Copepod 5 & Copepod 8 & Copepod 1 & Copepod 6\\
Locomotor classification & puller & puller& pusher & pusher\\
Prosome length ($\mathrm{m}$)	& 6.34 $\times$ 10$\mathrm{^{-4}}$ & 6.58 $\times$ 10$\mathrm{^{-4}}$ & 7.76 $\times$ 10$\mathrm{^{-4}}$  & 7.56 $\times$ 10$\mathrm{^{-4}}$\\
Prosome width ($\mathrm{m}$) & 4.46 $\times$  10$\mathrm{^{-4}}$ & 4.52 $\times$ 10$\mathrm{^{-4}}$ & 4.49 $\times$ 10$\mathrm{^{-4}}$ & 4.41 $\times$ 10$\mathrm{^{-4}}$\\
Body Re	& 0.93 & 2.96 & 7.37 & 6.02\\
Mean swimming speed (BL s$^{-1}$) & 2.37 & 6.99 & 12.54 & 10.78\\
Leg beat frequency ($\mathrm{s^{-1}}$) & 57.69\footnotemark[2] & 58.82\footnotemark[3] & 57.69\footnotemark[4] & 61.22\footnotemark[4]\\
Swimming direction (rad) & 1.555 & 1.529 & $-1.327$ & $-2.880$\\
Cephalic appendages beat motion & anterior to lateral & anterior to lateral & lateral to posterior & lateral to posterior\\
Relative anterior appendage start angle\footnotemark[1] & 0.28$\pm$0.01\footnotemark[2] & 0.23$\pm$0.01\footnotemark[3] & 0.47$\pm$0.04\footnotemark[4] & 0.57$\pm$0.02\footnotemark[4]\\
Relative anterior end angle\footnotemark[1] & 0.49\footnotemark[2] & 0.60\footnotemark[3] & 0.79\footnotemark[4] & 0.71\footnotemark[4]\\
\end{tabular}
\end{ruledtabular}
\footnotetext[1]{Relative to the swimming direction and normalized to 180$^{\circ}$.}
\footnotetext[2]{Calculated for 7 consecutive leg beats}
\footnotetext[3]{Calculated for 10 consecutive leg beats}
\footnotetext[4]{Calculated for 3 consecutive leg beats}
\end{table*}

\section{\label{sec:level1}Discussion}
Copepods and many other mesozooplankton species actively swim from the ocean surface down to several hundred meters deep and back up to avoid predation and feed \cite{Roe1972,Marshall1972}. In doing so, these organisms are subjected to external mechanical forces. For example, gravity was identified by Clarke \cite{Clarke1934} as a critical factor in negatively buoyant plankters like \textit{T. longicornis}, which tend to sink continuously. Previous works showed that calanoids – which lack internal gravity receptors – perform geotropic swimming in response to the retarding force of gravity, sensed via mechanoreceptors in their antennae \cite{Strickler1982,Bidder1929}. In general, copepods overcome the forces impeding forward swimming by using their relative weight to induce a stronger velocity gradient than downward-swimming copepods (see Fig.~\ref{fig:VeloVortPress}) \cite{Strickler1982,Emlet1985}. We found that this also contributes to stronger pressure gradients anteriorly.The literature reports how copepods often generate a funnel-shaped anterior flow prone to producing sub-ambient pressures \cite{Tiselius1990,Malkiel2003}. This phenomenon is often tied to feeding and hovering, whereby copepods generate a stronger anterior current, facilitating prey capture and manipulation \cite{Jiang2023,Gerritsen1977}. However, the link between pulling and pushing the water anteriorly and the swimming orientation has not been established in detail.\\

We found that the water just anterior to \textit{T. longicornus} will either be pushed forwards or pulled backwards during cruising depending on their orientation with respect to the gravitational acceleration (see Figs~\ref{fig:VeloVortPress},\ref{fig:Schematics}). This means that upwards swimmers utilize predominantly a `pull-based mechanism' while swimming whereas downwards swimming uses a  predominantly `push-based mechanism'. We present evidence that the onset of a strong pulling current works to drop the pressure directly in front of the copepods to enhance thrust when swimming upward (Fig.~\ref{fig:Forces}), likely as a compensatory mechanism to overcome gravity and cruise more economically than a similar but otherwise pushing copepod generating acceleration reaction forces anteriorly that contribute directly to drag. In contrast, downward-swimming `pushers' achieve greater cruising speeds. Following, we discuss how a subtle yet necessary shift of the leg movements modulates the structure of the near-field flow and consequently affects the propulsive forces of `pullers' and `pushers' in the context of vertical cruising.\\

Since the pulling mechanism is fundamental to feeding (i.e., increasing prey encounter rate and capture) and sensing \cite{Yen2013,Kiorboe2014,Kiorboe2013}, its role in locomotion, particularly during vertical relocation, has been overshadowed and remains to be explored. We found that by setting up a pulling current during upward swimming, copepods generate a strong sub-ambient pressure gradient directly in front of them (Fig.~\ref{fig:Forces}). This induces a pulling force on the body, which effectively generates net thrust during a portion of the leg recovery stroke, a phase during which both gravity and body drag otherwise resist the motion (Fig.~\ref{fig:Forces}(a),(e)). This is similar to the pull thrust mechanism described in fish, jellyfish, and ctenophores which promotes efficient thrust production for economical swimming \cite{Gemmell2015,Colin2020}. Note that this mechanism does not supplant the legs, and copepods still need to generate drag-based forces with their cephalic appendages. Nonetheless, copepods have evolved to be just the right size and swimming speed to take advantage of the inertial effects of the persistent anterior flow they induce to generate additional thrust during the generally drag-dominated leg recovery phase. Despite yielding slower swimming speeds, this strategy may help lower the COT compared with an upward `pusher' because it generates additional thrust rather than drag caused by anterior acceleration reaction forces. This is a worthwhile feature, considering these animals perform large-scale DVMs.\\

But what causes the contrasting near-field flow structure of `puller' and `pushers'? We present evidence for the ``anchor'' hypothesis suggested by \cite{Emlet1985} describing upward-swimming copepods using their relative weight to set up stronger velocity and pressure gradients. This was evidenced by the periodic oscillations in the forces, corresponding to forward acceleration phases (i.e., net thrust) and sinking due to the retarding effect of gravity (i.e., net drag). Crucial, however, is the difference in the initial position of the swimming legs during a beat effectively modulating the direction of the flow. `Pullers' performed `breaststroke' kinematics consisting of the cephalic appendages extending anteriorly and creating a vacuum when displaced laterally during the power stroke (Fig.~\ref{fig:VeloVortPress}). In contrast, `pushers' produce much weaker pressure gradients – with dominating positive pressures – because the initial lateral orientation of their swimming legs cannot induce a proper vacuum. An upside to this is the fact that the `pushers' are less likely to waste energy laterally since the normal component of the force produced by the legs is oriented more axially compared with `pullers'. This simple change in the leg kinematics is sufficient to modulate the near-field flow and promote conditions favorable to harnessing suction forces to generate more thrust. \\

Given the significant benefits of pulling, why do downward-swimming copepods not adopt it? They no longer need to generate forces to overcome gravity, and only drag opposes motion. In fact, the added effects of gravity contribute directly to increasing the overall swimming speed due to the added terminal velocity (see Table~\ref{tab:table1}) \cite{Clarke1934} and, at least partially, counter drag to produce thrust far more uniformly (Fig.~\ref{fig:Forces}(f)). Compared with upward copepods needing to overcome the effects of their weight, this undoubtedly requires less power \cite{Jiang2023} and thus potentially lowers the COT, thereby promoting efficient, fast cruising. This is central to their natural DVM behavior, for which they swim to greater depths. Note that gravity does not oppose motion during horizontal swimming, thus leaving drag as the only retarding force. As such, copepods are more likely to display the pushing behavior to swim faster. Our results show that even when a pronounced horizontal component exists, the swimming kinematics, induced flow, and forces are consistent with fast, downward `pushers'.\\

Copepods can still achieve fast swimming when not entirely assisted by gravity. However, whether behaving as pullers would be more advantageous in this context is unclear. Suction-based thrust might still be less efficient than cruising as a pusher with (or even without) the help of gravity. For this reason, pulling may be undesirable. Nonetheless, the pulling behavior demonstrates important benefits to enhance thrust when resisting gravity. One question remains unanswered, however: is an upward-swimming, pulling copepod more efficient than the same but otherwise pushing copepod? This may not be easily addressed experimentally, but computational fluid dynamics models can compare the energetics of upward- and downward-cruising pullers and pushers and quantify the impact this suction-based pulling behavior has in the context of DVM.


\bibliography{Copepod.bib}

\end{document}


\preprint{}

\title{Ups and downs: Copepods reverse the near-body flow to cruise in the water column}

\author{Nils B. Tack}
 \affiliation{ School of Engineering, Brown University, 345 Brook St., Providence, RI 02912, USA.}
\author{Sara Oliveira Santos}%
 \affiliation{ School of Engineering, Brown University, 345 Brook St., Providence, RI 02912, USA.}
\author{Brad J. Gemmell}
 \affiliation{Department of Integrative Biology, University of South Florida, 4202 East Fowler Ave., Tampa, FL 33620, USA.}
\author{Monica M. Wilhelmus}%
 \email{mmwilhelmus@brown.edu}
 \affiliation{ School of Engineering, Brown University, 345 Brook St., Providence, RI 02912, USA.}
 \email{mmwilhelmus@brown.edu}

\date{\today}

\pacs{}

\maketitle

\section{Swimming direction}
\begin{figure}[h!]
    \centering
    \includegraphics[scale=1]{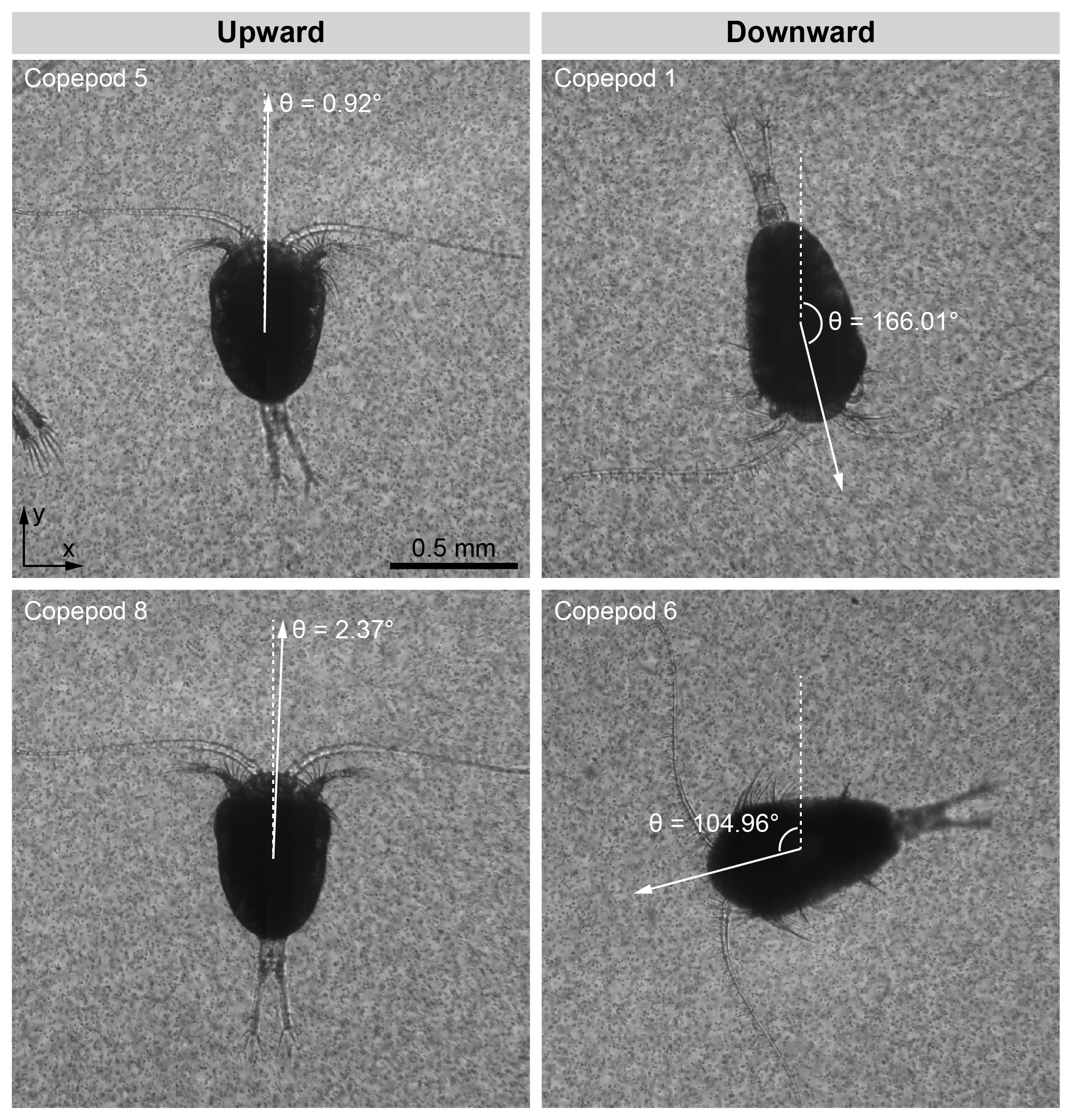}
    \caption{Swimming direction of the upward- and downward-swimming copepods. (Left panels) The heading of the upward swimmers was less than $\theta = 3^{\circ}$ offset from the vertical laboratory frame of reference (y). Copepod 1 swam downward nearly vertically (top right panel), while Copepod 6 had a strong horizontal component (bottom right panel). The white arrow indicates the mean swimming direction during the video sequence relative to y (dashed white line).}
    \label{fig:swimDirection}
\end{figure}

In total, we studied four freely swimming copepods; two animals swimming upward and two specimens swimming downward (Fig. S\ref{fig:swimDirection}). While the upward-swimming copepods swam nearly vertically ($u$ was less than 3$^{\circ}$ offset from the vertical laboratory frame of reference), one of the downward-swimming copepods displayed a strong horizontal component.

\section{Estimation of the sinking speed of \textit{T. longicornis}}
\noindent The gravitational settling velocity of an inactive copepod in water depends on the density and size (equivalent spherical diameter - ESD) of the animal as well as on the density and viscosity of the water. Copepod settling velocity was estimated from the mean density of 16 other calanoid copepod species as $\rho_{copepod}$ = 1066.6 kg m$^{3}$ (no density data exists for \textit{T. longicornis}) reported in the existing literature \cite{Mauchline1998}. The ESD was chosen as the width of the protosome of each of the four experimental copepods. The terminal sinking speed $\omega_s$ was calculated using a modified Stokes' law \cite{Ferguson2004,Elliot2010} for irregularly shaped particles as:
\begin{equation}
   \omega_s = \frac{RgD^{2}}{C_1 \nu+(0.75C_2 RgD^{3})^{0.5}}
\end{equation}

\noindent where $\omega_s$ is the particle (copepod) sinking velocity, $R$ is submerged specific gravity [$(\rho_{particle} - \rho_{fluid})/\rho_{fluid}$, where $\rho$ is density], $g$ is gravitational acceleration, $D$ is particle ESD, $\nu$ is fluid kinematic viscosity, and $C_1$ and $C_2$ are constants (24 and 1.2, respectively; as recommended for angular grains \cite{Ferguson2004}. Mean estimated $\omega_s$ = 2.68 $\pm$ 0.05 mm s$^{-1}$, equivalent to 3.82 $\pm$ 0.41 BL s$^{-1}$.

\section{Calculation of the depth of focus}
\noindent The depth of focus (DoF) of the present $\mu$-PIV system was evaluated according to \cite{Nakano2005}:

\begin{equation}
  DoF = \frac{n \ \lambda}{NA^{2}} + \frac{n \ e}{M \ NA}
\end{equation}

\noindent where $n$ is the refractive index of the fluid medium ($\approx$ 1.34 for our experiments), $\lambda$ is the source light wavelength ($\approx$ 0.7 $\mu$m for the longest wavelength for the white light used), $NA$ is the numerical aperture of the microscope objective (= 0.10), $M$ is the magnification of the microscope objective (= 4), and $e$ is the camera sensor pixel size (i.e., the smallest resolvable distance = 10 $\mu$m). The calculated DoF is $\approx$ 127 $\mu$m.

\section{Cephalic appendages orientation}
\begin{figure}[h!]
    \centering
    \includegraphics[scale=1]{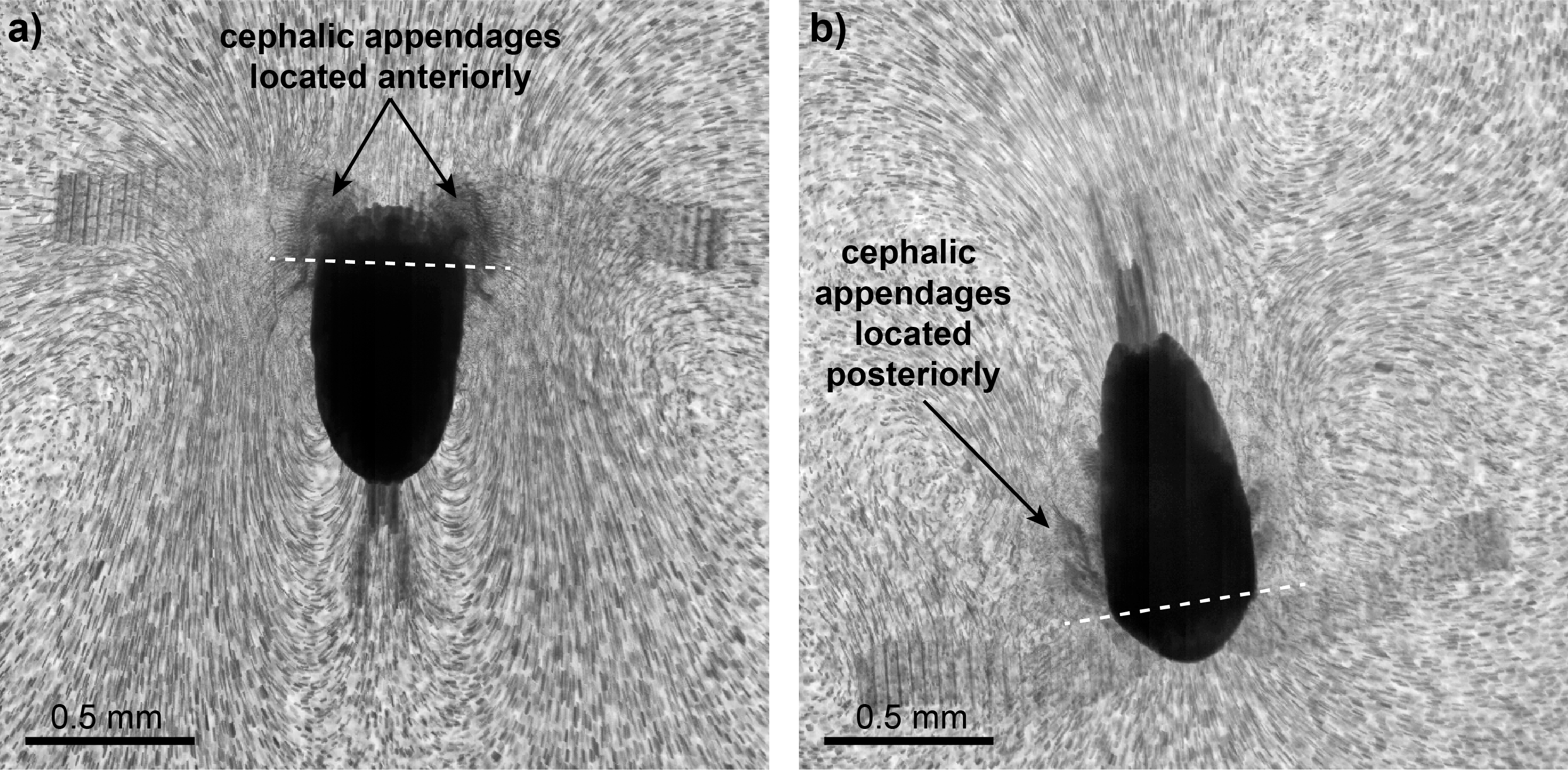}
    \caption{Image stacks of two representative copepods. (a) Upward swimmers maintained their legs anterior to the body during a leg beat cycle (copepod ID: 8; stack length = 35 frames = 0.0175 s). (b) Downward swimmers maintained their legs posteriorly to the body (Copepod ID: 1; stack length = 35 frames = 0.0175 s). The white dashed lines in both panels pass through the widest portion of the body in the last frame of the stack to serve as a visual reference to locate the legs relative to the body.}
    \label{fig:legPosition}
\end{figure}

 To quantify the differences in the leg kinematics during upward and downward swimming, we measured the angle of the primary appendages involved in producing the anterior and near-body flows relative to the swimming direction $U$ at the beginning and end of a power stroke. We identified the second antennae A2 (located anteriorly) as the primary appendages recruited during upward and downward swimming. Other cephalic appendages, including the mandibles (Md), first (Mx1) and second (Mx2) maxillae, and maxillipeds (Mxp), were also recruited but could not be tracked over time because they were obstructed by the prosome (due to the dorsoventral view). The angle formed relative to $U$ was normalized to 180$^{\circ}$. Normalized angles equal to 0.5 indicate a lateral orientation, while angles $<$0.5 and $>$0.5 indicate anterior and posterior orientations, respectively.\\
 The major appendage kinematics difference observed between upward and downward swimmers was the range of angles achieved by the cephalic appendages. Stacking sequential images of a complete leg beat cycle reveals the motions of the legs over time and their overall position relative to the body of the copepods (Fig. S\ref{fig:legPosition}). Upward swimmers performed ‘breaststroke’ movements by keeping the maxillipeds relatively anterior to the body (Fig. S\ref{fig:legPosition}a). Downward swimmers displayed ‘rowing’ motions by extending the legs laterally to the body (Fig. S\ref{fig:legPosition}b). 

\section{Calculating the pressure and force coefficients}
Pressure fields were used to calculate thrust and drag forces per height along 475 evenly sized segments around the body and appendages. The spacing between the boundary points corresponded to the spacing between velocity vectors (pressure data points) to limit the effects of interpolation. Force magnitude was calculated per unit depth (because PIV data were 2D) as the product of pressure, the length of each segment between points along the outline of the copepods, and the unit vector normal to each segment, giving units of Newtons per meter. Thrust and drag were computed from the axial component of the forces in the swimming direction. The force coefficient ($C_{F}$) for the push and pull components of thrust and drag were calculated as follows:

\begin{equation}
   C_F = \frac{F}{\frac{1}{2}\rho \ L \ U^{2}}
\end{equation}

\noindent where $F$ is the instantaneous axial force magnitude per unit depth (because PIV data were 2D) in N m$^{-1}$, $\rho$ is the density of 35 ppt water at 21$^\mathrm{o}$C (1024.238 kg m$^{-3}$), $U$ is the mean swimming speed of the copepod during the video sequence ($\mathrm{m \ s^{-1}}$), and $L$ is the body length of the copepods (the prosome only) in meter.\\

\noindent The pressure coefficient, ($C_{P}$) was calculated as follows:

\begin{equation}
   C_P = \frac{P}{\frac{1}{2}\rho \ U^{2}}
\end{equation}

\noindent where $P$ is the pressure (in Pa) calculated for each velocity vector, $\rho$ is the density of 35 ppt water at $21^\mathrm{o}$C ($1024.238 \ \mathrm{kg \ m^{-3}}$), $U$ is the mean swimming speed of the copepod during the video sequence ($\mathrm{m \ s^{-1}}$).

\section{Validating the cruising behavior}
Feeding and foraging are not mutually exclusive behaviors. Copepods can generate a feeding current that propels them through the water \cite{Yen2013,Kiorboe2014}. In the present study, the flows that propel the copepods are considered separate from those entraining particles (i.e., feeding). We evaluated the motions of the legs and the resulting flow to validate that our observations are more consistent with cruising rather than feeding. Copepods that produce a feeding current move the cephalic appendages in a manner that \cite{Strickler1984} describes as a ‘fling and clap’ motion. This mechanism is aimed at directing and trapping food particles. None of the copepods we observed – whether swimming upward or downward – displayed this behavior.
Additionally, \cite{strickler1982} shows that calanoid copepods tend to generate an anterior, double shear field when setting up a feeding current, which is absent in the cruising mode. As a result, the flows that entrain food particles organize into two vortices around the mouthparts, while the flows that propel the copepods form two larger medially opposing vortices around the cephalothorax (prosome) \cite{Cannon1928}. The flow features commonly described in feeding copepods were absent in the copepods we studied. Instead, we report flow features around the body that agree with the attributes of the cruising behavior. Therefore, we concluded that for all the cases presented in this study, including upward and downward animals, we investigated the cruising rather than feeding behavior (or combinations).\\

\begin{figure}[h!]
    \centering
     \includegraphics[width=1\textwidth]{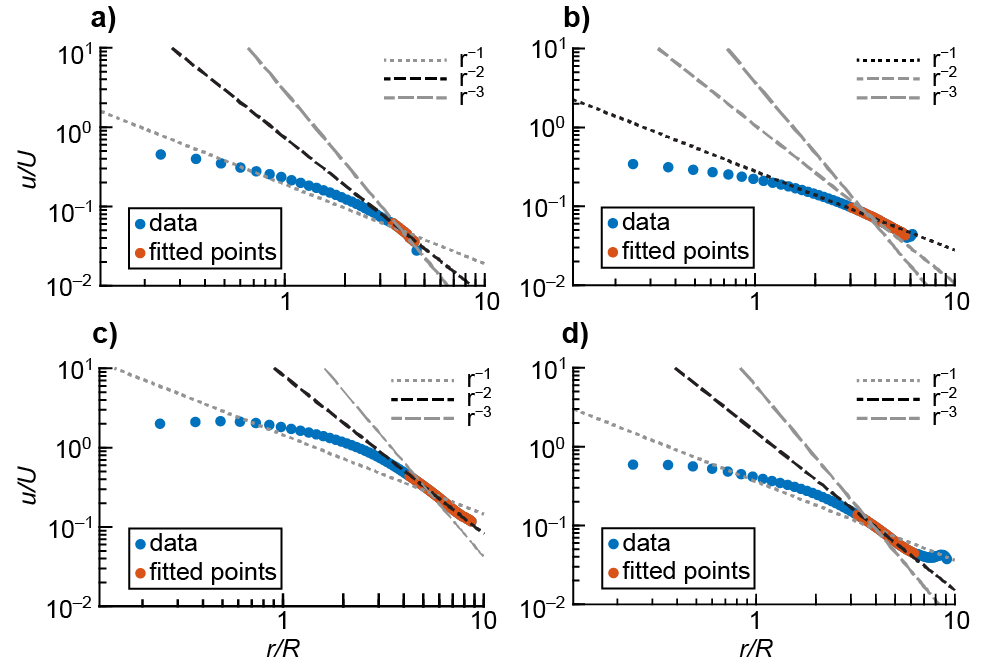}
    \caption{Velocity decay in front of the four copepods tested. Velocity decay is plotted as a function of the non-dimensional radius away from the copepod and the non-dimensional copepod velocity. (a-b) Downward swimming copepods (`pusher'). (c-d) Upward swimming copepods (`puller'). The flow in front of copepod 1, 5, and 8 (respectively a, c, and d) decays as $r^{-2}$, and the flow in front of copepod 6 (b) decays as $r^{-1}$. The orange data points were used to fit the leading order term $n$ of the velocity decay $r^{-n}$, starting at $r/R = 3$ to exclude near-body unsteady effects until the edge of the frame or nearest flow instability.}
\label{fig:VeloDecay}
\end{figure}

 The decay was measured sufficiently far away from the body ($r/R = 3$) to exclude near-body unsteady effects skewing attenuation estimates (\cite{drescher2010direct}) such that velocity decayed following a power relationship with the increasing distance from the body (Fig. S\ref{fig:VeloDecay}). The upper distance threshold from the body boundary was selected as the edge of the frame or the boundary of the nearest flow instability. Overall, the spatial attenuation (fluid velocity $u$ varied with distance $r$) of the flow disturbance produced by the downward and upward swimmers is consistent with other millimeter-sized zooplankton \cite{Kiorboe2014,Kiorboe2013,Kiorboe2010}. We found $u \propto r^{-2}$, which describes a routine cruising regime \cite{Kiorboe2014}. Faster attenuation is generally achieved in the order of $r^{-3}$ to $r^{-4}$ during fast swimming, such as predatory lunges and escapes. \cite{Kiorboe2014} have shown that plankton locomotion is optimized to reduce fluid disturbances during cruising, but simultaneous swimming while feeding has a greater disturbance, increasing their chance of being detected by a predator . The far-field attenuation for all copepods is between $r^{-1}$ and $r^{-2}$, comparable to an idealized stokeslet and stresslet model (Table S\ref{tab:table2}, Fig. S\ref{fig:VeloDecay}).\\

\begin{table}[h!]
\caption{\label{tab:table2} Velocity decay numerical fitting parameters. The power of the velocity decay in front of the copepod was estimated using a numerical fitting procedure that assessed the best match based on the goodness of fit. The fit was modeled as $ar^{-n}$. $a$ is the strength of the moment, the radial distance $r$ is normalized by the effective copepod radius $R$, and $n$ is the attenuation power coefficient.The dominant decay term corresponds to the highest coefficient of determination R$^{2}$.}

\begin{tabular}{cccc}
Copepod ID & Power \textit{n} & \textit{a} & R\textsuperscript{2}\\
\hline
1 & 1 & 0.19 & 0.71\\
 & 2 & 0.74 & 0.99\\
 & 3 & 2.84 & 0.85\\
\hline
5 & 1 & 1.45 & 0.77\\
 & 2 & 8.24 & 1.00\\
 & 3 & 42.08 & 0.75\\
\hline
6 & 1 & 0.28 & 0.95\\
 & 2 & 1.07 & 0.70\\
 & 3 & 3.79 & $-0.47$\\
\hline
8 & 1 & 0.36 & 0.81\\
 & 2 & 1.51 & 0.98\\
 & 3 & 5.78 & 0.62\\
 
\end{tabular}
\end{table}

\clearpage
\section{Reference}
\bibliography{Copepod.bib}